\documentclass[]{beilstein}

\usepackage{xspace}

\begin{document}

\title{Superconductor-insulator transition in capacitively coupled superconducting nanowires}
\author{Alex Latyshev}
\affiliation{I.E. Tamm Department of Theoretical Physics, P.N. Lebedev Physical Institute, 119991 Moscow, Russia}
\affiliation{National Research University Higher School of Economics, 101000 Moscow, Russia}
\author[1]{Andrew G. Semenov}
\affiliation{Department of Physics, Moscow Pedagogical State University, 119435 Moscow, Russia} 
\author*[2]{Andrei D. Zaikin}{andrei.zaikin@kit.edu}
\affiliation{Institute for Quantum Materials and Technologies, Karlsruhe Institute of Technology (KIT), 76021 Karlsruhe, Germany}

\nolinenumbers
\maketitle

\begin{abstract}
We investigate superconductor-insulator quantum phase transitions in ultrathin capacitively coupled superconducting nanowires
with proliferating quantum phase slips. We derive a set of coupled Berezinskii-Kosterlitz-Thouless-like renormalization group equations
demonstrating that interaction between quantum phase slips in one of the wires gets modified due to the effect of plasma modes
propagating in another wire. As a result, the superconductor-insulator phase transition in each of the wires is controlled not only by its own parameters but also by those of the neighboring wire as well as by mutual capacitance. We argue that superconducting nanowires with properly chosen parameters may turn insulating once they are brought sufficiently close to each other.
\end{abstract}


\section{Introduction}
Quantum fluctuations dominate the physics of superconducting nanowires at sufficiently low temperatures making their behavior markedly different from that of bulk superconductors \cite{book,AGZ,LV,Bezrbook}.  Many interesting properties of such nanowires are attributed to the effect of quantum phase slips (QPS) which correspond to fluctuation-induced local temporal suppression of the superconducting order parameter  inside the wire accompanied by the phase slippage process and quantum fluctuations of the voltage in the form of pulses. By applying a bias current one breaks the symmetry between positive and negative voltage pulses and, as a result,
 a superconducting nanowire acquires a non-vanishing electric resistance down to lowest temperatures \cite{ZGOZ,GZQPS}. This effect was directly observed in a number of experiments \cite{BT,Lau,Zgi08,liege}.  
 
Likewise, quantum phase slips in superconducting nanowires yield shot noise of the voltage \cite{SZ16} which originates from the process of quantum tunneling of magnetic flux quanta across the wire. One can also proceed beyond the voltage-voltage correlator and evaluate all cumulants of the voltage operator, thus deriving full counting statistics of quantum phase slips \cite{SZ19}. This theory enables one to obtain a complete description of superconducting fluctuations in such nanowires. Interesting QPS-related effects also occur in superconducting nanorings which can be employed, e.g., for possible realization of superconducting qubits \cite{MH}. Such effects were investigated theoretically \cite{SZ13} and observed experimentally \cite{As,Zhenya}. 

Each quantum phase slip generates sound-like plasma modes \cite{Mooij} which propagate along the wire and interact with other quantum phase slips. The exchange of such Mooij-Sch\"on plasmons produces logarithmic in space-time interaction between different QPS
which magnitude is controlled by the wire diameter (cross section) \cite{ZGOZ}. For sufficiently thick wires this interaction is strong and quantum phase slips are bound in close pairs. Accordingly, the (linear) resistance of such wires tends to zero at $T \to 0$, thus demonstrating a superconducting-like behavior in this limit. On the other hand, inter-QPS interaction in ultrathin wires is weak, quantum phase
slips are unbound and the superconducting phase fluctuates strongly along the wire. In this case the wire looses long scale superconducting properties, its total resistance remains non-zero and even tends to increase with decreasing temperature thus indicating an insulating behavior at $T \to 0$. At zero temperature the transition between these two types of behavior comes as a quantum phase transition (QPT) driven by the wire diameter \cite{ZGOZ}. Below we will also refer to this QPT as superconductor-insulator transition (SIT).

In this work we are going to show that this SIT can be substantially modified in a system of capacitively coupled superconducting nanowires even without any direct electric contact between them. In our previous work \cite{LSZ} we already elucidated some non-local QPS-related effects in such nanowires which yield non-equilibrium voltage fluctuations in the system which exhibit a non-trivial dependence on frequency and bias current. Here we will demonstrate that quantum fluctuations in one of the two wires effectively "add up" to those of another one, thereby shifting QPT in each of the wires in a way to increase the parameter range for the insulating phase. Qualitatively the same effect is expected to occur in a single superconducting nanowire that has the form of a meander frequently used in experiments.

\section{The model}
\begin{figure}
\caption{The systems under consideration: a) Two capacitively coupled superconducting nanowires and b) Superconducting nanowire in the form of a meander.}
\label{figure:1}
\includegraphics[width=16.0cm,keepaspectratio]{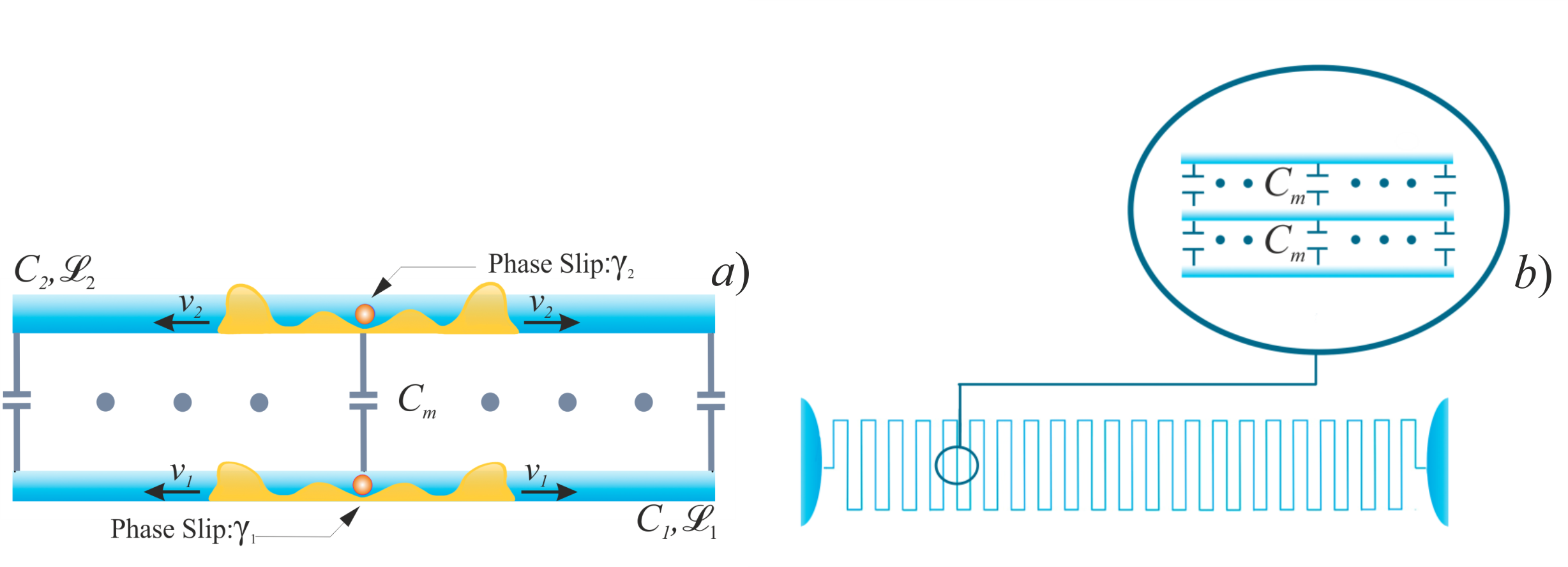}
\end{figure}

We first consider the system of two long parallel to each other superconducting nanowires, as it is schematically shown in Figure \ref{figure:1}a. 
The wires
are described by geometric capacitances $C_1$ and $C_2$ (per unit wire length) and kinetic inductances $\mathcal{L}_{1}$ and $\mathcal{L}_{2}$ (times length) effectively representing the two transmission lines. Capacitive coupling between these two nanowires is accounted for by the mutual capacitance $C_m$. The corresponding contribution to the system Hamiltonian that keeps track of both electric and magnetic energies in these coupled transmission lines reads  
\begin{equation}\label{eq4}
\hat{H}_{TL}=\frac{1}{2}\sum_{i,j=1,2}\int dx (\mathcal{L}^{-1}_{ij}\hat{\Phi}_{i}(x)\hat{\Phi}_{j}(x)+(1/\Phi^{2}_{0})C^{-1}_{ij}(\nabla \hat{\chi}_{i}(x)\nabla \hat{\chi}_{j}(x)),
\end{equation}
where $x$ is the coordinate along the wires, $\mathcal{L}_{ij}$ and $C_{ij}$ denote the matrix elements of the inductance and capacitance matrices
\begin{equation}\label{eq6}
\check{\mathcal{L}}=\left[\begin{array}{crl}
\mathcal{L}_{1} & 0\\
0 & \mathcal{L}_{2}\\\end{array}\right],   \quad \check{C}=\left[\begin{array}{crl}
C_{1} & C_{m}\\
C_{m} & C_{2}\\\end{array}\right]
\end{equation}
and $\Phi_0=\pi/e$ is the superconducting flux quantum. Note that for the sake of simplicity here and below we set Planck constant 
$\hbar$, speed of light $c$ and Boltzmann constant $k_B$ equal to unity.

The Hamiltonian (\ref{eq4}) is expressed in terms of the dual operators $\hat{\chi}(x)$ and $\hat{\Phi}(x)$ \cite{SZ13} which 
obey the canonical commutation relation 
\begin{equation}\label{eq2}
    [\hat{\Phi}(x),\hat{\chi}(x^{\prime})]=-i \Phi_{0}\delta(x-x^{\prime})
\end{equation} 
and are related to the charge density and the local phase operators, respectively $\hat{Q}(x)$ and $\hat{\varphi}(x)$, 
by means of the following equations
\begin{equation}\label{eq1}
    \hat{Q}(x)=\frac{1}{\Phi_{0}}\nabla\hat{\chi}(x), \;\;\;\;\;\; \hat{\varphi}=2e\int^{x}_{0}dx'\hat{\Phi}(x').
\end{equation}

Physically, $\hat{\Phi}_{i}(x)$ represents the magnetic flux operator, while the operator $\hat{\chi}_{i}(x)$ is proportional to that for the total charge $\hat{q}_{i}(x)$  that has passed through the point $x$ of the $i$-th wire up to the some time moment $t$, i.e. $\hat{q}_{i}(x)=-\hat{\chi}_{i}(x)/\Phi_{0}$.

Provided the wires are thick enough the low energy Hamiltonian in Eq. (\ref{eq4}) is sufficient. However, for thinner wires one should also account for the effect of quantum phase slips. The corresponding contribution to the total Hamiltonian for our system can be expressed in the form \cite{SZ13}
\begin{equation}\label{eq5}
\hat{H}_{QPS}=-\sum_{j=1,2}\gamma_{j} \int dx\cos(\hat{\chi}_{j}(x)),
\end{equation}
where 
\begin{equation}
\gamma_j \sim (g_{j \xi}\Delta/\xi)\exp (-ag_{j \xi}), \quad j=1,2
\label{gaQPS}
\end{equation}
denote the QPS amplitudes per unit wire length \cite{GZQPS}, $g_{j \xi} =R_q/R_{j \xi}$ is dimensionless conductance of the $j$-th wire segment of length equal to the superconducting coherence length $\xi$
(here and below $R_q =2\pi/e^2 \simeq 25.8$ K$\Omega$ is the quantum resistance unit and $R_{j \xi}$ is the normal state resistance of the corresponding wire segment), $\Delta$ is the superconducting order parameter and $a \sim 1$ is a numerical prefactor.  We also note that the Hamiltonian (\ref{eq5}) describes tunneling of the magnetic flux quantum $\Phi_0$ across the wire and can be viewed as a linear combination of creation ($e^{i\hat{\chi_{i}}}$) and annihilation ($e^{-i\hat{\chi_{i}}}$) operators for the flux quantum $\Phi_{0}$ .

It is obvious from Eq. (\ref{eq1}) that QPS events cause redistribution of charges inside the wire and generate pairs of voltage pulses moving simultaneously in the opposite directions (cf., Figure \ref{figure:1}a)
\begin{equation}\label{eq7}
\hat{V}_{i}(t)=1/\Phi_{0}\sum_{j=1,2}C^{-1}_{ij}(\nabla\hat{ \chi}_{j}(x_{1},t)-\nabla \hat{\chi}_{j}(x_{2},t)).
\end{equation}
Clearly, in the presence of capacitive coupling quantum phase slips in one of the wires also generate voltage pulses in another one.

To summarize the above considerations, the total Hamiltonian for our system is defined as a sum of the two terms in Eqs. (\ref{eq4}) and (\ref{eq5}),
\begin{equation}\label{eq3}
\hat{H}=\hat{H}_{TL}+\hat{H}_{QPS},
\end{equation} 
representing an effective sine-Gordon model that will be treated below.

\section{Quantum phase transitions: renormalization group analysis}
In order to quantitatively describe QPT in coupled superconducting wires we will employ the renormalization group (RG) analysis.
This approach is well developed and was successfully applied to a variety of problems in condensed matter theory, such as, e.g., 
the problem of weak Coulomb blockade in tunnel \cite{SZ90,GS,PZ,Zwerger} and non-tunnel \cite{KN,GZ04,BN} barriers between normal metals
or that of a dissipative phase transition in resistively shunted Josephson junctions \cite{SZ90,Albert,Blg84,GHM}. In the case of superconducting nanowires QPT was described \cite{ZGOZ} with the aid of RG equations equivalent to those initially developed for two-dimensional superconducting films \cite{BKT} which exhibit classical Berezinskii-Kosterlitz-Thouless (BKT) phase transition driven by temperature. In contrast, quantum SIT in quasi-one dimensional superconducting wires \cite{ZGOZ} with geometric capacitance $C$ and kinetic inductance ${\mathcal L}$ is controlled by the parameter  \cite{ZGOZ}
\begin{equation}
\lambda =\frac{R_q}{8}\sqrt{\frac{C}{\mathcal L}}
\label{la}
\end{equation}
 proportional to the square root of the wire cross section $s$.

It follows immediately from the analysis of Ref. \cite{ZGOZ} that provided the two superconducting wires  depicted in Figure \ref{figure:1}a are 
decoupled from each other, i.e. for $C_m \to 0$, one should expect two independent QPT to occur in these two wires respectively at $\lambda_1=2$ and at $\lambda_2=2$ where, according to Eq. (\ref{la}), we define $\lambda_{1,2} =(R_q/8)\sqrt{C_{1,2}/{\mathcal L}_{1,2}}$. The task at hand is to investigate the effect of capacitive coupling between the wires on these two QPT.

For this purpose let us express the grand partition function of our system ${\mathcal Z}={\rm Tr}\exp (-\hat{H}/T)$ in terms of the path 
integral
\begin{equation}
{\mathcal Z}=\int D\chi_1\int D\chi_2\exp (-S[\chi_1,\chi_2]),
\label{Z}
\end{equation}
where 
\begin{equation}\label{eq8}
    S=\frac{1}{2\Phi^{2}_{0}}\sum_{i,j=1,2}\int dxd\tau\Big(\xi\Delta\mathcal{L}_{ij}\partial_{\tau}\chi_{i}\partial_{\tau}\chi_{j}+\frac{1}{\xi\Delta}C^{-1}_{ij}\partial_{x}\chi_{i}\partial_{x}\chi_{j}\Big)-\sum_{i=1,2}y_{i}\int dxd\tau \cos\chi_{i}
\end{equation}
is the effective action corresponding to the Hamiltonian (\ref{eq3}) and 
\begin{equation}
y_{i}= \gamma_{i} \xi/\Delta \sim g_{j \xi}\exp (-ag_{j \xi}) \ll 1
\end{equation}
denote effective fugacity for the gas of quantum phase slips in the $i$-th wire. Note that, having in mind that the QPS core size in $x$- and $\tau$-directions is respectively $x_{0}\sim \xi$ and $\tau_{0}\sim \Delta^{-1}$, in Eq. (\ref{eq8}) for the sake of convenience we rescaled the spatial coordinate in units of $x_0$, i.e. $x \to x\xi $ and the time coordinate in units of $\tau_0$, i.e. $\tau \to \tau /\Delta$. 

In the spirit of Wilson's RG approach we routinely divide the $\chi$-variables into fast and slow components $\chi_{i}=\chi_{i}^{f}+\chi_{i}^{s}$, where
\begin{eqnarray}\label{eq9}
    \chi^{f}_{i}(x,\tau)=\int_{\Lambda<\omega^{2}+q^{2}<\Lambda+\delta\Lambda}\frac{d\omega dq}{2\pi}\chi_{\omega,q}e^{i\omega \tau+iqx},\;\;\;\chi^{s}_{i}(x,\tau)=\int_{\omega^{2}+q^{2}<\Lambda}\frac{d\omega dq}{2\pi}\chi_{\omega,q}e^{i\omega \tau+iqx}.
\end{eqnarray}

Setting $\delta\Lambda/\Lambda \ll 1$, expanding in the fast field components $\chi_{i}^{f}$ and integrating them out we proceed perturbatively in $y_{1,2}$ and observe that in order to account for the leading order corrections it is necessary to evaluate the matrix Green function at coincident points which reads
\begin{equation}
\check{G}^{f}(0,0)=\Phi^{2}_{0}\int \frac{d\omega dq}{(2\pi)^2}\Big(\xi\Delta\check{\mathcal L}\omega^2+\frac{1}{\xi\Delta}\check{C}^{-1}q^2\Big)^{-1}=2(\delta \Lambda/\Lambda)\check{\lambda},
\label{G00}
\end{equation}
where $\check{\lambda}=(R_q/8)\check{\mathcal{V}}\check{C}$ and $\check{\mathcal{V}}= (\check{C}\check{\mathcal L})^{-1/2}$ is the velocity matrix for plasmon modes propagating along the wires. The matrix $\check{\lambda}$ has the form
\begin{equation}\label{eq10}
    \check{
\lambda}=\frac{1}{\sqrt{\frac{1}{v^{2}_{1}}+\frac{1}{v^{2}_{2}}+\frac{2\sqrt{1-\frac{C_{m}^2}{C_1C_2}}}{v_{1}v_{2}}}}\left[\begin{array}{crl}
\lambda_1\left(\frac{1}{v_1}+\frac{\sqrt{1-\frac{C_{m}^2}{C_1C_2}}}{v_{2}}\right)& R_qC_{m}/8\\
R_qC_{m}/8& \lambda_2\left(\frac{1}{v_2}+\frac{\sqrt{1-\frac{C_{m}^2}{C_1C_2}}}{v_{1}}\right)\\\end{array}\right],
\end{equation}
where $v_i=1/\sqrt{C_i\mathcal{L}_i}$ is the velocity of the Mooij-Sch\"on modes in the $i$-th wire in the absence of capacitive coupling between the wires, i.e. for $C_m \to 0$.

Following the standard procedure \cite{BKT} and proceeding to bigger and bigger scales $\Lambda$, we eventually arrive at the following RG equations for the QPS fugacities $y_1$ and $y_2$: 
\begin{equation}\label{eq11}
    \frac{dy_{i}}{d\log \Lambda}=(2-\lambda_{ii})y_{i},\quad i=1,2,
\end{equation}
where $\lambda_{11}$ and $\lambda_{22}$ are diagonal elements of the matrix $\check{\lambda}$ (\ref{eq10}). Note that here we restrict our RG analysis to the lowest  order in $y_{1,2}$ which is sufficient for our purposes. As long as one keeps only the linear in $y_{1,2}$ terms in the RG equations all other parameters of our problem, e.g., $\lambda_{ii}$, remain unrenormalized. 

As it can be observed from Eqs. (\ref{eq11}), our system exhibits two BKT-like QPT at $\lambda_{11}=2$ and $\lambda_{22}=2$. In the limit $C_m \to 0$ the wires are independent from each other,  $\lambda_{11(22)}\to\lambda_{1(2)}$ and these QPT obviously reduce to that predicted in Ref. \cite{ZGOZ}. However, for non-zero capacitive coupling between the wires  the two QPT occur at the values of $\lambda_{1,2}$ exceeding 2. For the first wire the corresponding phase transition point is fixed by the condition
\begin{equation}
\lambda_1=2\frac{\sqrt{1+\frac{v^2_1}{v^2_2}+2\frac{v_1}{v_2}\sqrt{1-\frac{C_m^2}{C_1C_2}}}}{1+\frac{v_1}{v_2}\sqrt{1-\frac{C_m^2}{C_1C_2}}}.
\label{QPT}
\end{equation}
The same condition for the second wire is obtained from Eq. (\ref{QPT}) by interchanging the indices $1 \leftrightarrow 2$.

The above results allow to conclude that in the presence of capacitive coupling SIT in both wires occurs at larger values of $\lambda_{1,2}$ than in the absence of such coupling. In other words, quantum fluctuations in one of these wires effectively decrease superconducting properties of the other one. 

It follows from Eq. (\ref{QPT}) that the magnitude of such mutual influence depends on the ratio of the plasmon velocities in the two wires $v_1/v_2$ and on the strength of the capacitive coupling controlled by $C_m$. Provided the wire cross sections $s_1$ and $s_2$ differ strongly the plasmon velocities $v_i \propto \sqrt{s_i}$ also differ considerably. Assume, for instance, that the first wire is much thinner than the second one. In this limit we have $v_1 \ll v_2$ and, hence, the QPT condition (\ref{QPT}) in the first wire remains almost unaffected for any capacitive coupling strength. If, on the contrary, the first wire is much thicker than the second one, then one has $v_1 \gg v_2$ and the condition (\ref{QPT}) reduces to $\lambda_1\simeq 2/\sqrt{1-C_m^2/(C_1C_2)}$ demonstrating that the critical value $\lambda_1$ can exceed 2 considerably for sufficiently large values $C_m$. 

\begin{figure}
\caption{a) Critical surfaces corresponding to SIT at $\lambda_{11}=2$ and $\lambda_{22}$=2.
b) Phase diagram for two capacitively coupled superconducting nanowires with $\lambda_{1}=2.01$ and $\lambda_{2}=2.03$.
Both curves $\lambda_{11}(C_m)$ and $\lambda_{22}(C_m)$ decrease and cross the critical line $\lambda_c=2$ with increasing mutual capacitance $C_m$.}
\label{figure:2}
\includegraphics[width=16.0cm,keepaspectratio]{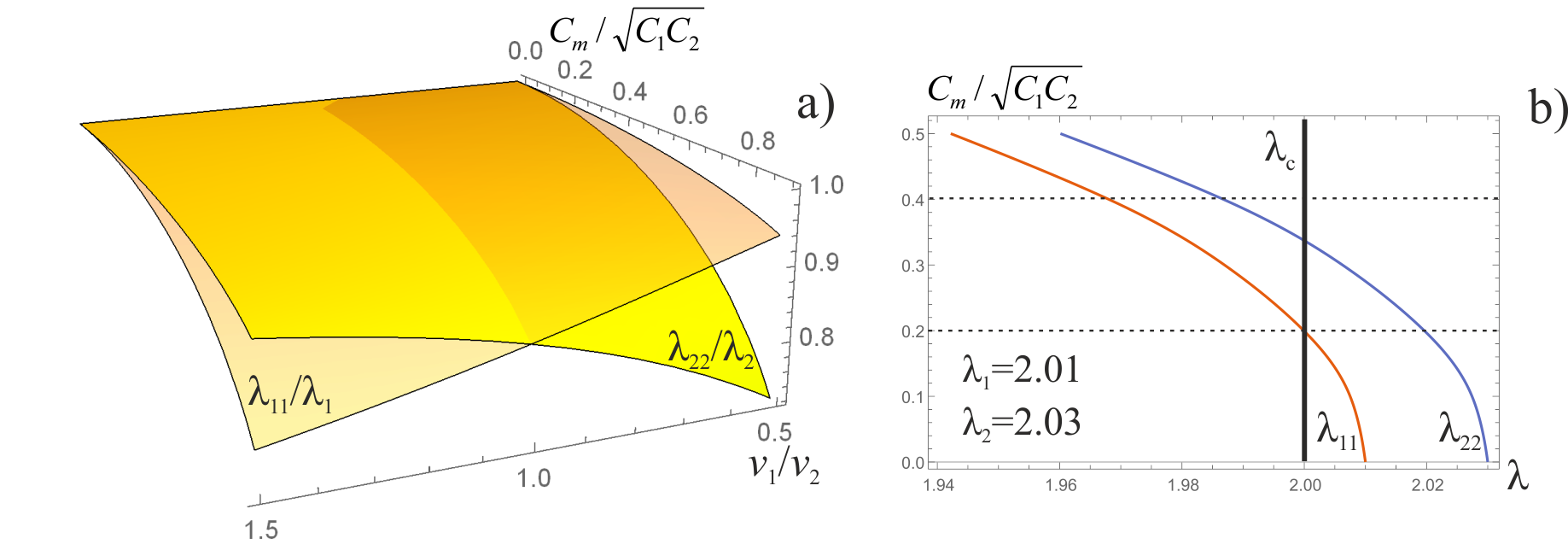}
\end{figure}

It is obvious that the strength of capacitive coupling depends on the distance between the wires. At large distances this coupling is negligible $C_m \to 0$, whereas as the wires get closer to each other the value $C_m$ increases and, hence, their mutual influence increases as well. Let us choose the wire parameters in such a way that for $C_m =0$ both these wires remain in the superconducting phase being relatively close to SIT. In this case the parameters $\lambda_1$ and $\lambda_2$ should be just slightly bigger than 2. Moving the wires closer to each other we "turn on" capacitive coupling between them, thus, decreasing both values $\lambda_1$ and $\lambda_2$ below 2. As a result, two superconducting wires become insulating as soon as they are brought sufficiently close to each other. This remarkable physical phenomenon is illustrated by the phase diagram in Figure \ref{figure:2}b.

In order to complete this part of our analysis we point out that transport properties can be investigated in exactly the same manner 
as it was done, e.g., in Ref. \cite{ZGOZ} in the case of a single nanowire. Generalization of the technique \cite{ZGOZ} to the case of two capacitively coupled superconducting nanowires is straightforward. For a linear resistance of the $i$-th wire $R_{i}(T)$ and for $\lambda_{ii} >2$ (or for any $\lambda_{ii}$ at sufficiently high temperatures) we obtain
\begin{equation}\label{eq17}
    R_{i}(T)\propto \gamma^{2}_{i} T^{2\lambda_{ii}-3}, \quad i=1,2.
\end{equation}

\section{Extension to other geometries}

The effects discussed here can be observed in a variety of structures involving superconducting nanowires. For instance, superconducting nanowires in the form of a meander (see Figure \ref{figure:1}b) are frequently employed in experiments, see, e.g., Ref. \cite{Gre}. In this case different segments of the wire are parallel to each other being close enough to develop electromagnetic coupling. Having in mind the above analysis one expects that the wire of such a geometry would be "less superconducting" than the same wire that has the form of a straight line.

For an illustration, let us mimic the behavior of the wire depicted in Figure \ref{figure:1}b by considering three identical parallel to each other 
capacitively coupled superconducting nanowires. For simplicity we will assume the nearest neighbor interaction, i.e. the second (central) nanowire is coupled to both the first and the third nanowires via the mutual capacitance $C_m$ whereas the latter two are decoupled from each other. We again assume that the wires are thin enough and quantum phase slips may proliferate in each of these wires. 

Quantum properties of this system are described by the same effective action (\ref{eq8}) where the inductance and capacitance 
matrices now take the form
\begin{eqnarray}
\check{\mathcal{L}} =\left[\begin{array}{crl}
\mathcal{L} & 0 & 0\\
0 &\mathcal{L}& 0\\
0& 0& \mathcal{L}\\\end{array}\right], \quad 
\check{C}=\left[\begin{array}{crl}
C & C_{m} & 0\\
C_{m} & C & C_{m}\\
0& C_{m}& C\\\end{array}\right],\end{eqnarray}
and the summation runs over the indices $i,j=1,2,3$. Proceeding along the same lines as in the previous section we again arrive at 
Eq. (\ref{G00}), where the diagonal elements of the matrix $\check{\lambda}$ now read
\begin{eqnarray}
\label{22}
\lambda_{22}=\frac{\lambda}{2}\left(\sqrt{1-\sqrt{2}\frac{C_{m}}{C}} + \sqrt{1+\sqrt{2}\frac{C_{m}}{C}}\right),\\
\lambda_{11}=\lambda_{33}= \frac{\lambda}{2}\left(1+\frac{1}{2}\left(\sqrt{1-\sqrt{2}\frac{C_{m}}{C}} + \sqrt{1+\sqrt{2}\frac{C_{m}}{C}}\right)\right)
\label{1133}
\end{eqnarray}
and the QPS interaction parameter $\lambda$ is defined in Eq. (\ref{la}). We again arrive at the RG equations of the form (\ref{eq11}) (now with $i=1,2,3$). Being combined with Eqs. (\ref{22}), (\ref{1133}) these RG equations demonstrate that in the presence of capacitive coupling SET occur at $\lambda_{ii}=2$ implying $\lambda > 2$ for each of the three wires. This observation is fully consistent with our previous results derived for two coupled nanowires.

Furthermore, the RG equation (\ref{eq11}) with $i=2$ combined with Eq. (\ref{22}) also describes the effect of interacting quantum phase slips and QPT in the wire having the form of a meander (Figure \ref{figure:1}b). In this case, within the approximation of the nearest neighbor capacitive interaction between the wire segments QPT occurs at
\begin{equation}
\lambda= \frac{4}{\sqrt{1-\sqrt{2}\frac{C_{m}}{C}} + \sqrt{1+\sqrt{2}\frac{C_{m}}{C}}},
\end{equation}
i.e. the critical value of the parameter $\lambda$ exceeds 2 as soon as the mutual capacitance $C_m$ differs from zero. As it is clear from Eqs. (\ref{22}), (\ref{1133}), the approximation of the nearest neighbor interaction appears to be well justified in the limit $C_m \ll C$. For stronger interactions with $C_m \sim C$ this approximation most likely becomes insufficient for a quantitative analysis. However, on a qualitative level our key observations should hold also in this case: A nanowire in the form of a straight line with $\lambda$ slightly exceeding the critical value 2 should demonstrate superconducting-like behavior with $R(T) \propto T^{2\lambda -3}$ \cite{ZGOZ} whereas the wire with exactly the same parameters may turn insulating provided it has the form of a meander with capacitive coupling between its segments.


\section{Results and Discussion}
We have analyzed the effect of quantum fluctuations in capacitively coupled superconducting nanowires. We have demonstrated that plasma modes propagating in one such nanowire play the role of an effective quantum environment for another one modifying the logarithmic interaction between quantum phase slips in this wire. As a result, the superconductor-insulator quantum phase transition gets shifted in a way to increase the parameter range for the insulating phase. Hence, superconducting nanowires may turn insulating provided they are brought close enough to each other. It would be interesting to observe this effect in forthcoming experiments with superconducting nanowires.



\begin{funding}
We acknowledge partial support by RFBR Grant No. 18-02-00586.
AGS acknowledges support by the Russian Science Foundation (project No. 19-72-10101).
\end{funding}


\begin{thebibliography}{}
\bibitem{book} Zaikin, A.D.; Golubev, D.S. {\it Dissipative Quantum Mechanics of Nanostructures: Electron Transport, Fluctuations and Interactions}; Jenny Stanford: Singapore, 2019. doi:10.1201/9780429298233
\bibitem{AGZ} Arutyunov, K.Y.; Golubev, D.S.; Zaikin, A.D. {\it Phys. Rep.} {\bf 2008}, 464, 1-70. doi:10.1016/j.physrep.2008.04.009
\bibitem{LV} Larkin, A.I.; Varlamov, A.A. {\it Theory of fluctuations in superconductors}; Clarendon: Oxford, 2005. doi:10.1093/acprof:oso/9780198528159.001.0001
\bibitem{Bezrbook} Bezryadin, A. {\it Superconductivity in Nanowires}; Wiley-VCH: Weinheim, 2013. ISBN 978-3-527-40832-0
\bibitem{ZGOZ} Zaikin, A.D.; Golubev, D.S.; van Otterlo, A.; Zimanyi, G.T. {\it Phys. Rev. Lett.} {\bf 1997}, 78, 1552-1555. doi:10.1103/PhysRevLett.78.1552
\bibitem{GZQPS} Golubev, D.S.; Zaikin, A.D. {\it Phys. Rev. B} {\bf 2001}, 64, 014504. doi:10.1103/PhysRevB.64.014504
\bibitem{BT} Bezryadin, A.; Lau, C.N.; Tinkham, M. {\it Nature} {\bf 2000}, 464, 971-973. doi:10.1038/35010060
\bibitem{Lau} Lau, C.N.; Markovic, N.; Bockrath, M.; Bezryadin, A.; Tinkham, M. {\it Phys. Rev. Lett.} {\bf 2001}, 87, 217003. doi:10.1103/PhysRevLett.87.217003
\bibitem{Zgi08} Zgirski, M.; Riikonen, K. P.; Touboltsev, V.; Arutyunov, K.Y. {\it Phys. Rev. B} {\bf 2008}, 77, 054508. doi:10.1103/PhysRevB.77.054508
\bibitem{liege} Baumans, X.D.A.; Cerbu, D.; Adami, O-A.; Zharinov, V.S.; Verellen, N.; Papari, G.; Scheerder, J.E.; Zhang, G.; Moshchalkov, V.V.; Silhanek, A.V.; Van de Vondel, J.  {\it Nat. Commun.} \textbf{2016}, 7, 10560. doi:10.1038/ncomms10560
\bibitem{SZ16} Semenov, A.G.; Zaikin, A.D. Phys. Rev. B {\bf 2016}, 94, 014512. doi:10.1103/PhysRevB.94.014512
\bibitem{SZ19} Semenov, A.G.; Zaikin, A.D. Phys. Rev. B {\bf 2019}, 99, 094516. doi:10.1103/PhysRevB.99.094516
\bibitem{MH}  Mooij, J.E.; Harmans, C.J.P.M. {\it New J. Phys.} {\bf 2005},  7, 219. doi:10.1088/1367-2630/7/1/219
\bibitem{SZ13} Semenov, A.G.; Zaikin, A.D. Phys. Rev. B {\bf 2013}, 88, 054505. doi:10.1103/PhysRevB.88.054505
\bibitem{As} Astafiev, O.V.; Ioffe, L.B.; Kafanov, S.; Pashkin, Y.A.; Arutyunov, K.Y.; Shahar, D.; Cohen, O.; Tsai, J.S. {\it Nature} {\bf 2012}, 484, 355-358. doi:10.1038/nature10930
\bibitem{Zhenya}  De Graaf, S.E.;  Skacel, S.T.;  H\"onigl-Decrinis, T.; Shaikhaidarov, R.; Rotzinger, H.; Linzen, S.;
Ziegler, M.; H\"ubner, U.;  Meyer, H.-G.; Antonov, V.; Il'ichev, E.;  Ustinov, A.V.; Tzalenchuk, A.Ya.; Astafiev, O.V. {\it Nat. Phys.} {\bf 2018}, 14, 590-594. doi:10.1038/s41567-018-0097-9
\bibitem{Mooij} Mooij, J.E.; Sch\"on, G. {\it Phys. Rev. Lett.} {\bf 1985}, 55, 114-117. doi:10.1103/PhysRevLett.55.114
\bibitem{LSZ} Latyshev, A.; Semenov, A.G.; Zaikin, A.D. {\it J. Supercond. Nov. Magn.} {\bf 2020}, 33, 2329-2334. doi: 10.1007/s10948-019-05402-3
\bibitem{SZ90} Sch\"on, G.; Zaikin, A.D. {\it Phys. Rep.} {\bf 1990}, 198, 237-412. doi:10.1016/0370-1573(90)90156-v
\bibitem{GS} Guinea, F.; Sch\"on, G. {\it EPL} {\bf 1986}, 1, 585-594. doi:10.1209/0295-5075/1/11/007
\bibitem{PZ} Panyukov, S.V.; Zaikin, A.D. {\it Phys. Rev. Lett.} {\bf 1991}, 67, 3168-3171. doi:10.1103/PhysRevLett.67.3168
\bibitem{Zwerger} Hofstetter, W.; Zwerger, W. {\it Phys. Rev. Lett.} {\bf 1997}, 78, 3737-3740. doi:10.1103/PhysRevLett.78.3737
\bibitem{KN} Kindermann, M.; Nazarov, Yu.V. {\it Phys. Rev. Lett.} {\bf 2003}, 91, 136802. doi:10.1103/PhysRevLett.91.136802
\bibitem{GZ04} Golubev, D.S.; Zaikin, A.D.  {\it Phys. Rev. B} {\bf 2004}, 69, 075318. doi:10.1103/PhysRevB.69.075318
\bibitem{BN} Bagrets, D.A.; Nazarov, Yu.V. {\it Phys. Rev. Lett.} {\bf 2005}, 94, 056801. doi:10.1103/PhysRevLett.94.056801
\bibitem{Albert} Schmid, A. {\it Phys. Rev. Lett.} {\bf 1983}, 51, 1506-1509. doi:10.1103/PhysRevLett.51.1506
\bibitem{Blg84} Bulgadaev, S.A. {\it JETP Lett.} {\bf 1984}, 39, 315-319. WoS:A1984TH36800008.
\bibitem{GHM} Guinea, F.; Hakim, V.; Muramatsu, A. {\it Phys. Rev. Lett.} {\bf 1985}, 54, 263-266. doi:10.1103/PhysRevLett.54.263
\bibitem{BKT}  Kosterlitz, J.M. {\it J. Phys. C} {\bf 1974}, 7, 1046-1060. doi: 10.1088/0022-3719/7/6/005
\bibitem{Gre} Delacour, C; Pannetier, B.; Villegier, J.-C.; Bouchiat, V. {\it Nano Lett.} {\bf 2012}, 12, 3501-3506. doi: 10.1021/nl3010397
\end{thebibliography}

\end{document}